\documentclass[reprint,
superscriptaddress,
amsmath,
amssymb,
aps,
prb,
floatfix,
english
]{revtex4-2}

\usepackage{bibunits}
\usepackage[unicode=true, colorlinks=true, citecolor={blue!80!black}, urlcolor={blue!50!black}, linkcolor = {blue!80!black}]{hyperref}
\defaultbibliography{references}
\defaultbibliographystyle{apsrev4-2}
\usepackage{graphicx}
\usepackage{svg}
\usepackage{dcolumn}
\usepackage{siunitx}
\usepackage[T1]{fontenc}
\usepackage[utf8]{inputenc}
\usepackage{bbm}

\begin{document}
\title{Quantum theory of Bloch oscillations in a resistively shunted transmon
}

\author{Vladislav D.~Kurilovich}
\thanks{
Present address: Google Research, Mountain View, CA, USA.\\
Email: vladislav.kurilovich@gmail.com}
\affiliation{Department of Physics, Yale University, New Haven, CT 06520, USA}
\author{Benjamin Remez}
\affiliation{Department of Physics, Yale University, New Haven, CT 06520, USA}
\author{Leonid I.~Glazman}
\affiliation{Department of Physics, Yale University, New Haven, CT 06520, USA}

\begin{abstract}
     A transmon qubit embedded in a high-impedance environment acts in a way dual to a conventional Josephson junction. In analogy to the AC Josephson effect, biasing of the transmon by a direct current leads to the oscillations of voltage across it. These oscillations are known as the Bloch oscillations.
     We find the Bloch oscillations spectrum, and show that the zero-point fluctuations of charge make it broad-band. Despite having a broad-band spectrum, Bloch oscillations can be brought in resonance with an external microwave radiation. The resonances lead to steps in the voltage-current relation, which are dual to the conventional Shapiro steps. We find how the shape of the steps depends on the environment impedance $R$, parameters of the transmon, and the microwave amplitude. The Bloch oscillations rely on the insulating state of the transmon which is realized at impedances exceeding the Schmid transition point, $R > R_Q = h / (2e)^2$.
\end{abstract}
\maketitle
\section{Introduction}
A coherent charge propagation across a tunnel junction between superconductors gives rise to a celebrated DC and AC Josephson effects. A spectacular manifestation of the latter one is monochromatic current oscillations at frequency $2eV_J/\hbar$ in a junction biased by voltage $V_J$ \cite{josephson1962}. The Josephson effect is based on the continuous flow of the superconducting condensate across the junction. 
{The notion of the continuous flow of condensate is in tension with the Coulomb blockade phenomenon stemming from the charge discreteness.} A picture of charge-$2e$ Cooper pairs tunneling across the junction one after another have lead one to the prediction of Bloch oscillations \cite{likharev1985}, a phenomenon dual to the AC Josephson effect. Application of a direct current $I_J$ to a small Josephson junction leads to the accumulation of the displacement charge across the junction until it reaches $2e$, 
at which point a Cooper pair tunnels. The process repeats itself with the {angular} frequency $\Omega_J = 2\pi I_J / 2e$, resulting in the oscillations of the voltage across the current-biased junction; these are the Bloch oscillations.

The presence of Bloch oscillations relies on the {\it insulating} state of the junction. Indeed, the charge transferred through the junction is a variable conjugate to the superconducting phase difference. By uncertainty relation, having a well-defined, discrete transferred charge requires the phase variable to be delocalized. The latter condition corresponds to an insulating (as opposed to a superconducting) state of the junction. 


According to the Schmid transition paradigm \cite{schmid1983}, the Josephson junction becomes insulating if the impedance of its electromagnetic environment $R$ exceeds the resistance quantum $R_Q = h / (2e)^2$.
In fact, verification of this superconductor-to-insulator transition proved to be a challenging task by itself.
The voltage-current characteristics of the resistively-shunted junctions were measured in this context in Ref.~\cite{sonin1999}. 
The phase diagram extracted from these measurements deviated noticeably from the theory prediction.
A number of recent works goes as far as to contest the notion of the superconductor-to-insulator transition in a Josephson junction altogether~\cite{joyez2020, pekola2022, yuto2022, joyez2023}.
In opposition to these works, microwave experiments with Josephson junction arrays do seem compatible with the Schmid transition prediction \cite{kuzmin2023}.  

The Schmid transition controversy leads one to question if Bloch oscillations of a Josephson junction exist at all. 
Recent experimental works give evidence for this effect by reporting observation of dual Shapiro steps~\cite{shaikhaidarov2022, crescini2023, kaap2024} {(as well as of a related effect \cite{lotkhov2024})}. 
Dual Shapiro---or Bloch-Shapiro---steps \cite{likharev1985} arise from synchronization of Bloch oscillations with the external microwave radiation applied to the junction. The steps appear in the voltage across the junction and are centered at $I_J = n\cdot 2e\,\omega_{\rm ac} / 2\pi$, where $\omega_{\rm ac}$ is the microwave frequency and $n \in \mathbb{Z}$.

Admittedly, the quantization of the measured steps  
is much less precise than that of the conventional Shapiro steps in superconducting junctions centered around $V_J = n \cdot \hbar \omega_{\rm ac}/ 2e$ \cite{shapiro1968}.
The quantization of the latter is in fact so perfect that it is used as a metrological {\it voltage} standard. 
What are the fundamental limitations on the sharpness of the Bloch-Shapiro steps? 
This question---although important for experiments as well as for developing a metrological {\it current} standard---has not been answered.
Little is known \cite{averin1990} about an intimately related question of how monochromatic the frequency spectrum of Bloch oscillations is. 

To address these questions, we consider  a transmon qubit (i.e., a large Josephson junction) coupled to an Ohmic electromagnetic environment such as a resistor or a transmission line. {The advantage of the transmon is in a large gap separating its lowest Bloch band from the higher-energy excitations.}
The dynamics of the transmon within its lowest Bloch band is governed by the boundary sine-Gordon model.
Previously, application of this model allowed one to reveal signatures of the Schmid transition in microwave response of the transmon \cite{houzet2023, burshtein2023}. 
Here, we use it to find the manifestation of Bloch oscillations in the transport properties of the junction, as well as in its radiation spectrum. 

Our theory gives specific predictions for the voltage-current characteristics,
and points out the features in them indicating the insulating state of the junction. 
We also elucidate circuit parameters controlling the presence and sharpness of Bloch-Shapiro steps. 

\section{Model}
We consider a transmon qubit embedded in a high-impedance electromagnetic environment. 
A transmon is a variety of a Josephson junction in which the Josephson energy $E_J$ exceeds the charging energy $E_C$.
The condition $E_J \gg E_C$ guarantees that the energy spectrum of the transmon consists of well-separated charge bands. 
The separation between the bands suppresses the Landau-Zener tunneling allowing one to focus on the qubit dynamics within a single, isolated band. 
These are the optimal conditions for the observation of Bloch oscillations.

\begin{figure}[t]
  \begin{center}\includegraphics[scale = 1]{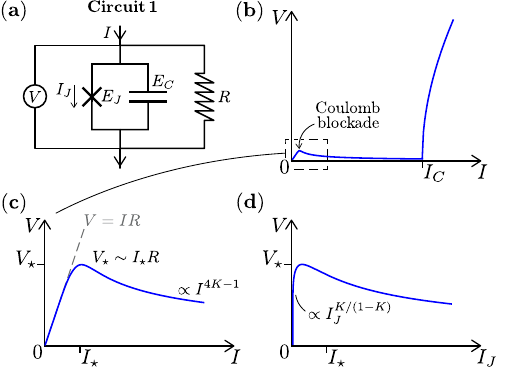}
    \caption{(a) In {\bf circuit 1}, a transmon is shunted by resistor $R$ and biased by current $I$. (b) If $R$ exceeds the resistance quantum $R_Q$, then the transmon ($E_J \gg E_C$) acts as an insulator. The insulating behavior is confined to a narrow domain of the $VI$-relation, $I \lesssim I_\star$ [see Eqs.~\eqref{eq:phase_slip} and \eqref{eq:I_star}]. For higher currents, $V(I)$ is similar 
    to that of a conventional superconducting junction. (c) Sketch of the low-current part of $V(I)$ [we use Eqs.~\eqref{eq:VI} and \eqref{eq:deltaV} with $K = 1/8$ to produce the plot]. (d)~The Coulomb blockade results in a sharp increase of $V$ with the current through the junction $I_J$, see Eq.~\eqref{eq:V_vs_IJ_blockade} and the discussion following it.}
    \label{fig:circuit1}
  \end{center}
\end{figure}

\begin{figure}[t]
  \begin{center}\includegraphics[scale = 1]{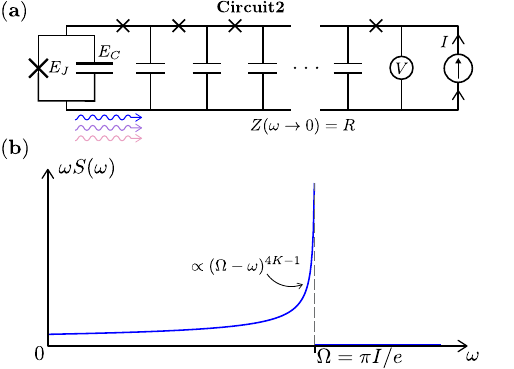}
    \caption{(a) In {\bf circuit 2}, a transmon is coupled to the Josephson junction array with a low-frequency impedance $R$. The voltage $V$ is measured at the same port at which the current $I$ is supplied. Because of the current-induced Bloch oscillations, transmon emits waves into the array. {(b) Power spectrum of the emitted waves $\omega S(\omega)$ is broad-band. Bloch oscillation frequency $\Omega = \pi I / e$ sets a characteristic frequency of an emitted photon, as well as the position of a threshold in the dependence of $\omega S(\omega)$ on $\omega$. The curve is produced using Eq.~\eqref{eq:S_omega} for $K = 1/8$.}}
\label{fig:circuit2}
  \end{center}
\end{figure}

Specifically, we consider two circuits depicted in Figs.~\ref{fig:circuit1} and~\ref{fig:circuit2}. In Fig.~\ref{fig:circuit1}, a transmon is shunted by a resistor $R$, and biased by an external current source $I(t)$. In Fig.~\ref{fig:circuit2}, a transmon is galvanically coupled to a transmission line
comprised of Josephson junctions); the current bias is supplied via the same 
line.
In $\omega \rightarrow 0$ limit, the impedance of the line $Z(\omega)$ approaches a constant $Z(0) \equiv R$ \footnote{Here we assume that the limit $L \rightarrow \infty$ (where $L$ is the length of the transmission line) is taken before $\omega \rightarrow 0$. We address the finite-size effects in a forthcoming work~\cite{Mismatch}. }. The latter feature allows us to model the resistor of Fig.~\ref{fig:circuit1} as a semi-infinite transmission line as well \cite{caldeiraleggett1981}.
At this level, the only difference between the two circuits is whether the junction and the transmission line are connected in parallel [Fig.~\ref{fig:circuit1}] or in series [Fig.~\ref{fig:circuit2}]. 

Both circuits can be described by the Hamiltonian of the form
\begin{equation}\label{eq:hamiltonian}
    H = H_J + H_R,
\end{equation}
where terms $H_J$ and $H_R$ correspond to the transmon and the transmission line, respectively. The Hamiltonian of the transmon is given by
\begin{equation}\label{eq:junction}
    H_J = 4E_C (N - n)^2 - E_J \cos \varphi.
\end{equation}
The first term is the electrostatic energy; the charging energy $E_C = e^2 / 2C$ is determined by the junction capacitance $C$. $N$ is the operator of the number of Cooper pairs transferred through the junction and  $n$ is the ``displacement'' charge imposed on the junction by the remaining circuit. 
The second term in Eq.~\eqref{eq:junction} is the Josephson coupling. The phase difference operator $\varphi$ is canonically conjugate to the Cooper pair number $N$: $[N, \varphi] = -i$.

The degrees of freedom of the electromagnetic environment are described by the term $H_R$ in Eq.~\eqref{eq:hamiltonian}. This term is given by 
\begin{equation}\label{eq:H_R}
    H_R = \int_0^\infty dx\,\frac{\hbar v}{2\pi}\left[\frac{1}{K}(\partial_{x}\theta)^{2}+K(\partial_{x}\phi)^{2}\right],
\end{equation}
where $v$ is the wave velocity in the transmission line and $K$ is a dimensionless parameter characterizing the line impedance~$R$,
\begin{equation}
    K = \frac{R_Q}{2R},\quad\quad\quad R_Q = \frac{h}{4e^2}.
\end{equation}
Hamiltonian \eqref{eq:H_R} is expressed in terms of the charge displacement field $\theta(x)$ and the phase field $\phi(x)$. 
The field $\theta(x)$ is related to the charge density in the transmission line, $\rho(x) = 2e\,\partial_x \theta / \pi$. 
The field $\phi(x)$ is canonically conjugate to the charge density, i.e., it satisfies the following commutation relation
\begin{equation}
    [\partial_x \theta(x), \phi(x^\prime)] = i\pi \delta (x - x^\prime).
\end{equation}
The boundary value of the phase field gives the phase difference across the junction, $\varphi \equiv \phi(x = 0)$. The continuous-field Hamiltonian \eqref{eq:H_R} adequately describes the junction array [Fig.~\ref{fig:circuit2}] as long as the relevant wavelengths exceed the array period.

For an isolated transmon, the displacement charge $n$ is a $c$-number. This makes the Hamiltonian \eqref{eq:junction} identical in form to the Hamiltonian of a quantum-mechanical particle moving in a periodic potential $\propto\!-\cos\varphi$. Parameter $n$ plays the role of the quasi-momentum in this analogy. Similarly to the spectrum of a particle in the periodic potential, the spectrum of the transmon consists of ``Bloch'' energy bands $E_j(n)$ which depend periodically on $n$ (with a period $1$). 
For $E_J \gg E_C$, the lowest Bloch band is given by
\begin{equation}\label{eq:lowest_band}
    E_0(n) = - \lambda \cos(2\pi n).
\end{equation}
The bandwidth $\lambda$ is determined by the amplitude of the phase slip at the junction (i.e., tunneling between two equivalent minima of the $-\cos\varphi$ ``potential''); it is exponentially small in $E_J / E_C \gg 1$ \cite{koch2007}:
\begin{equation}\label{eq:phase_slip}
    \lambda = E_C\,2^5\sqrt{\frac{2}{\pi}} \Bigl(\frac{E_J}{2E_C}\Bigr)^{3/4} e^{-\sqrt{8E_J / E_C}}.
\end{equation}
The lowest band is separated from the higher bands by an energy gap. The magnitude of the gap $E_{\rm gap} = \hbar \omega_{\rm Q} \gg \lambda$ is set by the plasma frequency of the junction,
\begin{equation}
\omega_{\rm Q} = \sqrt{8E_J E_C}/\hbar.
\end{equation}

The coupling of the transmon to the environment and current bias make the displacement charge $n$ a dynamical variable. There are two contributions to $n$:
\begin{equation}\label{eq:displacement}
    n = - \theta(x = 0) / \pi + {\cal N}.
\end{equation}
The first contribution is the charge transferred from one side of the junction to the other through the transmission line. The second contribution is a $c$-number that describes the current bias $I(t)$ applied to the circuit, 
$\dot{{\cal N}}(t) = I(t) / 2e$, see Fig.~\ref{fig:circuit1}. In the case of the circuit depicted in Fig.~\ref{fig:circuit2}, one may also use Eq.~\eqref{eq:displacement}  by including the current bias in the definition of $\theta(x)$ and modifying the boundary condition for $\theta(x)$ accordingly.

If $n$ changes in time slowly (on a scale set by the plasma frequency $\omega_{\rm Q}$), then interband transitions of transmon can be neglected; the system follows the variations of $n$ adiabatically.
One can then describe the dynamics of the circuit with the help of an effective Hamiltonian obtained by projecting the Hamiltonian \eqref{eq:junction} onto the lowest Bloch band. Using Eqs.~\eqref{eq:lowest_band} and \eqref{eq:displacement}, we obtain
\begin{equation}\label{eq:H_J_proj}
    H_J = -\lambda \cos(2\theta(0) - 2\pi {\cal N}),
\end{equation}
where $\lambda$ is given by Eq.~\eqref{eq:phase_slip}. 

Under the DC bias, the component of the displacement charge ${\cal N}$ grows linearly in time, ${\cal N} = I t / 2e$. As a result,  $H_J$ {\it oscillates} in time giving rise to Bloch oscillations of voltage across the junction. 

We now use the boundary sine-Gordon model defined by equations~\eqref{eq:hamiltonian}, \eqref{eq:H_R}, and \eqref{eq:H_J_proj} to find the spectrum of Bloch oscillations [Sec.~\ref{sec:spectrum}] and their manifestations in the transport properties of the transmon [Secs.~\ref{sec:voltage_current} and \ref{sec:shapiro}]. {Below, we use units with $\hbar = 1$.}




\section{Voltage-current characteristic\label{sec:voltage_current}}

First, we evaluate the DC voltage-current relation $V(I)$ for the two circuits of Figs.~1 and 2. 
As we will see, the character of $V(I)$ depends on the comparison between impedance $R$ and the resistance quantum $R_Q$. The transmon acts as a superconductor at $R < R_Q$, and has the traits of an insulator for $R > R_Q$. 


To start with, we consider the circuit in Fig.~\ref{fig:circuit1} and compute $V(I)$ perturbatively in the phase slip amplitude $\lambda$.
Because of the Bloch oscillations, the transmon biased by current $I$ acts as a source of waves emitting energy into the transmission line. 
The power $P$ dissipated via the wave emission can also be attributed to a DC voltage drop $V$ across the transmon:
\begin{equation}\label{eq:PeqIV}
    P = I V.
\end{equation}
We can evaluate the power $P$ using Fermi's golden rule and thus get $V(I)$.
Denoting the initial and final states of the circuit as $|i\rangle$ and $|f\rangle$, respectively, we obtain at $T = 0$:
\begin{align}\label{eq:P_start}
    P = 2\pi \Omega \sum_f |\langle f|\lambda\, e^{2i\theta(0)} / 2|i\rangle|^2 \delta(E_f - E_i - \Omega).
\end{align}
We introduced here
\begin{equation}\label{eq:Omega_to_I}
    \Omega =2\pi I / 2e,
\end{equation}
which is the frequency of oscillations in the Hamiltonian $H_J$ [cf.~Eq.~\eqref{eq:H_J_proj}]. With the considered accuracy, the majority of the supplied current flows through the junction, $I_J \approx I$, so $\Omega$ coincides with the Bloch oscillations frequency $\Omega_J = 2\pi I_J / 2e$. 
{We re-write the right-hand side of Eq.~\eqref{eq:P_start} in terms of the correlation function ${\cal C}_\theta(\Omega)$ of the boundary displacement field,}
\begin{align}\label{eq:P_to_C}
    P = \frac{\lambda^2 \Omega}{4}\,&{\cal C}_\theta(\Omega),\notag\\
    &{\cal C}_\theta(\Omega) = \int dt e^{i\Omega t}\langle e^{-2i\theta(0,t)} e^{2i\theta(0,0)}\rangle.
\end{align}
The averaging here is performed over the ground state of the circuit unperturbed by the phase slips, and $\theta(0,t) = e^{iH_R t} \theta(0) e^{-iH_R t}$, where $H_R$ is given by Eq.~\eqref{eq:H_R}. For concreteness, we assumed $I=2e\Omega / 2\pi > 0$.

The correlation function of a free boson theory is well-known \cite{weiss1985}:
\begin{equation}\label{eq:corr}
{\cal C_\theta}(\Omega) = \frac{2\pi}{\Gamma(4K)}\frac{1}{\Omega} \Bigl(\frac{\Omega}{\omega_{\rm Q}}\Bigr)^{4K} e^{-\Omega / \omega_{\rm Q}}\cdot \Theta(\Omega),
\end{equation}
where $\Theta(x)$ is the step function.
The exponent of the power-law factor is determined by impedance $R$ ({recall} 
that $K = R_Q / 2R$). The UV cutoff of the low-energy theory is set by the plasma frequency of the junction $\omega_{\rm Q}$~\cite{houzet2023}. 

By combining Eqs.~\eqref{eq:P_to_C} and \eqref{eq:corr} with Eq.~\eqref{eq:PeqIV}, and relating $\Omega$ to the current $I$ via Eq.~\eqref{eq:Omega_to_I}, we find that the $VI$--relation is a power-law:
\begin{equation}\label{eq:VI}
    V = \frac{\pi\lambda^2}{2\Gamma(4K)}\frac{1}{I} \left(\frac{\pi|I|}{e\omega_{\rm Q}}\right)^{4K},
\end{equation}
where we assumed $|I| \ll 2e \omega_{\rm Q}/ 2\pi$.

Equation~\eqref{eq:VI} reveals a transition between insulating and superconducting phases of the transmon as a function of $R$~\cite{schmid1983}. To see this, we assess the effective resistance $R_{\rm eff} = V / I$. From Eq.~\eqref{eq:VI} it follows that
\begin{equation}
    R_{\rm eff} \propto |I|^{4K - 2}.
\end{equation}
For $K > 1/2$ (i.e., at the low impedance, $R < R_Q$), the effective resistance vanishes at $I \rightarrow 0$; the transmon acts as a superconductor. The situation reverses for high impedance, $K < 1/2$ ($R > R_Q$). In this case, $R_{\rm eff}$ \textit{diverges} at low biases. 
The divergence reflects the onset of Coulomb blockade of transport across the junction.
The Coulomb interaction-driven superconductor-to-insulator transition happening across $R = R_Q$ is the Schmid transition. 

The low-bias divergence of $R_{\rm eff}$ at $R > R_Q$ signals a breakdown of perturbation theory in $\lambda$.  We can estimate the value of the bias $I_{\star}$ at which the breakdown happens by setting $R_{\rm eff}(I_\star) = R$. This condition gives
\begin{equation}\label{eq:I_star}
    I_\star = \frac{e\omega_{\rm Q}}{\pi} \left(\sqrt{\frac{2K}{\Gamma(4K)}}\frac{\pi\lambda}{\omega_{\rm Q}}\right)^{1/(1 - 2K)}.
\end{equation} 
The obtained result for $V(I)$ [Eq.~\eqref{eq:VI}] applies only at $I \gg I_\star$. The character of the $V(I)$-dependence changes qualitatively at low currents. 
At $I \ll I_\star$, the transmon acts as an almost perfect insulator. Therefore, the majority of the supplied current flows through the resistor [Fig.~\ref{fig:circuit1}], and only its small part $I_J \ll I$ flows through the junction. It means that the $V(I)$-relation of the circuit is close to the Ohmic one, 
\begin{equation}\label{eq:almost_Ohmic}
    V = I R - \delta V,\quad\quad\quad \delta V = I_J R.
\end{equation}
Here $I_J \equiv I_J(I)$ is a function of the total current $I$.
The Coulomb blockade effect is revealed the most directly in the $V(I_J)$-relation for the Josephson junction. Full blockade corresponds to a jump in $V(I_J)$. The jump is smeared by quantum fluctuations.
We now quantify the smearing by finding $I_J(I)$ and $V(I_J)$.

In the Coulomb blockade regime, the charge transport through the junction happens via rare events of a single Cooper pair tunneling.  The latter are described by an effective Hamiltonian dual to Eq.~\eqref{eq:H_J_proj}
\begin{equation}\label{eq:cos_phi}
    H_J = \tilde{\lambda} \cos (\varphi - 2e V t), 
\end{equation}
where $e^{i\varphi}$ is an operator that transfers a Cooper pair across the junction. 
In the leading-order approximation, the voltage drop is $V \approx I R$. 
The relation between the tunneling amplitude $\tilde{\lambda}$ in Eq.~\eqref{eq:cos_phi} and the phase slip amplitude $\lambda$ was derived in Ref.~\cite{fendley1995}:
\begin{equation}\label{eq:tilde_lambda}
    \tilde{\lambda} = {\frac{\omega_{\rm Q}}{\pi} \frac{\Gamma(1/2K)}{2 K}\Bigl(\frac{1}{2K \Gamma(2K)} \frac{\pi \lambda}{\omega_{\rm Q}}\Bigr)^{-1/2K}}.
\end{equation}
The Hamiltonian \eqref{eq:cos_phi} captures the least-irrelevant transport processes (in the renormalization group sense).

We now use Eq.~\eqref{eq:cos_phi} to find current $I_J$ through the Josephson junction. Employing Fermi's golden rule, we find at $T = 0$
\begin{equation}
    I_J = 2e\cdot 2\pi\sum_f|\langle f|\tilde{\lambda}e^{i\varphi}/2|i\rangle|^2 \delta(E_f - E_i - 2eV),
\end{equation}
where $|i\rangle$ is the ground state of the circuit, and $|f\rangle$ is the final state. Performing the summation over the final states, we express $I_J$ in terms of the correlation function of the phase difference~$\varphi$:  
\begin{align}\label{eq:I_J}
    I_J = \frac{e \tilde{\lambda}^2}{2}&{\cal C}_\varphi (2eV),\notag\\
    &{\cal C}_\varphi (2eV) = \int dt e^{i2eVt}\bigl\langle e^{-i\varphi(t)}e^{i\varphi(0)}\bigr\rangle.
\end{align}
Up to a replacement $4K \rightarrow 1 / K$, the latter correlation function coincides with that of the displacement charge [cf.~Eq.~\eqref{eq:corr}]. Therefore, we find
\begin{equation}\label{eq:I_J_vs_V}
    I_J = \frac{\pi \tilde{\lambda}^2}{2\Gamma(1/K)}\frac{1}{V}\Bigl(\frac{2e|V|}{{{\omega}_{\rm Q}}}\Bigr)^{1/K}.
\end{equation}
As a result, we obtain for the $V(I_J)$-relation of the junction:
\begin{equation}\label{eq:V_vs_IJ_blockade}
    V = \mathrm{sign}\,I_J\cdot\frac{\omega_{\rm Q}}{2e} \left(\frac{\Gamma(1/K)\omega_{\rm Q}}{\pi \tilde{\lambda}^2} \frac{|I_J|}{e}\right)^{K / (1-K)}.
\end{equation}
The Coulomb blockade results in a sharp increase of the voltage with $I_J$. At $I_J \sim I_\star$, the voltage reaches the crossover value $V_\star \sim I_\star R$. The Coulomb blockade of the Josephson junction breaks down at higher currents in which case  $I_J \approx I$; the $V(I_J)$-relation is given by Eq.~\eqref{eq:VI} with $I \rightarrow I_J$. Note that at a sufficiently high impedance, $K < 1/4$, $V(I_J)$ becomes a non-monotonic function, see Fig.~\ref{fig:circuit1}(d). A non-monotonic $V(I_J)$ was observed, e.g., in Ref.~\cite{watanabe2001}.

We can also use Eq.~\eqref{eq:I_J_vs_V} to quantify the deviation of the full $V(I)$-relation of the circuit from Ohm's law. 
Approximating  $V \approx I R$ in the right hand side of  Eq.~\eqref{eq:I_J_vs_V}, we obtain for the deviation $\delta V = I_J R$ in Eq.~\eqref{eq:almost_Ohmic}:
\begin{equation}\label{eq:deltaV}
    \delta V =  \frac{\pi \tilde{\lambda}^2}{2\Gamma(1/K)}\frac{1}{(2K)^{1/K}}\frac{1}{I }\Bigl(\frac{\pi|I|}{e\omega_{\rm Q}} \Bigr)^{1/K}.
\end{equation}
It remains small as long as $I \ll I_\star$.

At $K \ll 1$, the crossover scale $I_\star \sim e K \lambda$. In this limit, quantum fluctuations of charge can be neglected, and the circuit can be described by classical equations of motion. The solution of these equations yields the entire $V(I_J)$ dependence \cite{likharev1985}:
\begin{equation}\label{eq:classical_V_I_J}
    V =  I_JR \,\bigl[\sqrt{1 + (4eK \lambda / I_J)^2} - 1\bigr].
\end{equation}
The $V(I)$-relation of the full circuit is
\begin{equation}\label{eq:K0limit1}
V = 
\begin{cases}
    IR,&|I| < 4e K \lambda,\\
    IR\,\bigl[1 - \sqrt{1 - (4eK\lambda/I)^2}\bigr],&|I| > 4eK\lambda.
\end{cases}
\end{equation}
It agrees with the found asymptotes of $V(I)$ at $K \ll 1$.


{Next, we briefly address the $VI$-relation in the circuit of Fig.~\ref{fig:circuit2}. 
The principal difference between this circuit and the one in Fig.~\ref{fig:circuit1} is that in the former circuit {\it all} of the supplied current $I$ {\it has} to flow through the junction. 
This feature becomes important in the Coulomb blockade regime, $I \lesssim I_\star$. There, the need to overcome the blockade results in a sharp increase of the voltage drop $V$ with $I$.
The specific dependence $V(I)$ can be obtained by replacing $I_J \rightarrow I$ in Eq.~\eqref{eq:V_vs_IJ_blockade}; this yields $V \propto I^{K / (1 - K)}$.
On the other hand, at current $I \gg I_\star$, the difference between the two circuits becomes inconsequential and $V(I)$ is given by Eq.~\eqref{eq:VI}, same as for the circuit 1.} {Figure 1(d) with the replacement $I_J\to I$ illustrates the overall form of the $VI$-relation.}

\section{Spectrum of radiation emitted by Bloch oscillations\label{sec:spectrum}}
Due to the Bloch oscillations, the current-biased transmon acts as an antenna emitting waves into the transmission line [see Fig.~\ref{fig:circuit2}(a)]. 
Here we find the spectrum of the emitted photons.
We will see that the zero-point charge fluctuations make the spectrum non-monochromatic.


To the second order in $\lambda$, number of photons $S(\omega)d\omega$ emitted per unit time in a frequency range $[\omega, \omega + d\omega]$ can be evaluated with the help of Fermi's golden rule. A calculation similar to the one yielding Eq.~\eqref{eq:P_to_C} gives
\begin{equation}
    S(\omega) = \frac{4K}{\omega} \frac{\lambda^2}{4}{\cal C}_\theta(\Omega - \omega),
\end{equation}
where ${\cal C}_\theta(\Omega)$ is the displacement charge correlation function. Using Eq.~\eqref{eq:corr}, we find
\begin{equation}\label{eq:S_omega}
    S(\omega) = \frac{1}{\omega} \frac{2K\pi\lambda^2}{\omega_{\rm Q}\Gamma(4K)} \left(\frac{\Omega - \omega}{\omega_{\rm Q}}\right)^{4K-1} \cdot\Theta(\Omega - \omega).
\end{equation}
The Bloch oscillation frequency $\Omega = 2\pi I / 2e$ sets 
{the characteristic frequency of the emitted photons \footnote{We note that for circuit (b) all of the current flows through the junction, $I_J = I$, and therefore the Bloch oscillation frequency $\Omega_J$ coincides with $\Omega$ of Eq.~\eqref{eq:Omega_to_I}.}, but the radiation spectrum is broadened by the dynamics of charge quantum fluctuations.}
The profile of the spectral density $S(\omega)$ depends on the line impedance $R = R_Q / 2 K$.
For a high impedance line, $K < 1 / 4$, it becomes divergent at $\omega = \Omega$, {see Fig.~\ref{fig:circuit2}(b). }
In spite of the divergence, the spectrum remains broad: the entire interval $[0, \Omega]$ contributes to the total emitted power \footnote{This is similar to how the fluctuations destroy the M{\"o}ssbauer effect in one dimension.}. The spectrum becomes monochromatic only in a singular limit $K \rightarrow 0$.
Result \eqref{eq:S_omega} is valid at high biases $I \gg I_\star$. The decrease of $I$ towards $I_\star$ further broadens the spectrum by introducing additional thresholds $\propto \Theta(n \Omega - \omega)$ at multiples of~$\Omega$. 

\section{Bloch-Shapiro steps\label{sec:shapiro}}

\begin{figure}[t]
  \begin{center}\includegraphics[scale = 1]{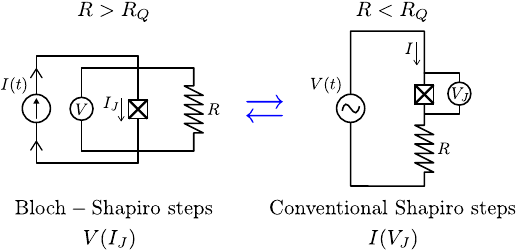}
    \caption{Bloch-Shapiro steps develop in the $V(I_J)$ dependence at the current values $ne\omega/\pi$ associated with the frequency of the ac component of the source current $I(t)$. These steps occuring in a circuit with $R>R_Q$ are dual to the conventional Shapiro steps in $I(V_J)$ of a superconducting junction in series with resistor $R < R_Q$. 
    {Boxes around the Josephson junctions indicate the junction capacitances.}
    }
    \label{fig:duality}
  \end{center}
\end{figure}

Bloch oscillations can be synchronized with the external microwave radiation.
Synchronization occurs whenever the Bloch oscillation frequency is an integer multiple of the microwave frequency $\omega_{\rm ac}$. For circuit (a), this condition is $\Omega_J = n\omega_{\rm ac}$, where $\Omega_J$ is set by the current $I_J$ flowing through the junction, $\Omega_J = 2\pi I_J / 2e$.
Synchronization leads to the steps in the $V(I_J)$-relation of the junction centered at $I_J = n \cdot 2e \omega_{\rm ac} / 2\pi$.
The steps in $V(I_J)$ in a shunted transmon are dual to conventional Shapiro steps in $I(V_J)$ of a Josephson junction in series with a resistor [Fig.~\ref{fig:duality}].
Here we find the dependence of the steps shape on the impedance $R$, properties of the transmon, and power of the microwave radiation.


To demonstrate how synchronization affects the $V(I_J)$-relation, we start with a Heisenberg equation of motion for the variable $\theta(0, t)$ following directly from Eqs.~\eqref{eq:H_R} and \eqref{eq:H_J_proj} \cite{likharev1985}:
\begin{align}\label{eq:shaps0}
    \frac{d\theta(0, t)}{dt}= \frac{d\theta^{(0)}(0, t)}{dt}
    -2\pi K\lambda\sin\left(2\theta(0, t)- 2\pi {\cal N}(t)\right).
\end{align}
The left hand side is the operator of current flowing through the resistor, with charge measured in units of $2e / \pi$. On the right hand side, $\theta^{(0)}(0, t)$ is the free field operator (i.e., the field operator unperturbed by the phase slips). It encodes the effect of quantum fluctuations on the dynamics of $\theta(0, t)$.
In the presence of microwave radiation, the bias supplied to the circuit is given by 
\begin{equation} \label{eq:alpha}
2\pi {\cal N}(t) =  \Omega t + \alpha \sin\omega_{\rm ac} t,
\end{equation}
where the first term describes the DC component, $\Omega = 2\pi I / 2e$, and $\alpha\cdot\omega_{\rm ac}$ is the amplitude of the microwave-induced AC component. 

Bloch-Shapiro steps are narrow if the microwave frequency is sufficiently high, $\omega_{\rm ac} \gg 2\pi I_\star / 2e$. Steps occur near the currents through the circuit $I$ close to values $n\cdot 2e \omega_{\rm ac}/2\pi$ set by the microwave radiation frequency, $|I - n \cdot 2e \omega_{\rm ac} / 2\pi| \ll 2e \omega_{\rm ac} / 2\pi$.
In what follows, we assume that the two conditions above are satisfied. To find the shape of the $n$-th step, we expand the oscillation in Eq.~\eqref{eq:shaps0} in the Fourier series, and single out the near-resonant harmonic:
\begin{align}\label{eq:shaps_2}
    &\frac{d\theta(0, t)}{dt}= \frac{d\theta^{(0)}(0, t)}{dt}\notag\\
    &\quad-2\pi K \sum_{p \neq n} J_p(\alpha) \lambda \sin(2\theta(0, t) - (\Omega - p \omega_{\rm ac})t)\notag\\
    &\quad-2\pi K J_n(\alpha)\lambda\sin\left(2\theta(0, t)- (\Omega - n \omega_{\rm ac})t\right).
\end{align}
Here $J_p(\alpha)$ is the Bessel function of order $p$. 
The off-resonant terms (second line) lead to the emission of high-frequency waves with $\omega \sim \omega_{\rm ac}$. As was discussed in Sec.~\ref{sec:voltage_current}, the radiation contributes to the DC voltage drop across the transmon. To the second order in $\lambda$, this effect can be captured in Eq.~\eqref{eq:shaps_2} by replacing the second line with a constant term $(\pi / 2e) V_n / R$, where
\begin{equation}
    V_n = \sum_{p \neq n} \frac{\pi \lambda^2}{2\Gamma(4K)}\frac{J^2_p(\alpha)}{I - p\tfrac{2e\omega_{\rm ac}}{2\pi}}\Bigl(\frac{\pi|I - p \tfrac{2e\omega_{\rm ac}}{2\pi}|}{e\omega_{\rm Q}}\Bigr)^{4K}. 
\end{equation}
see Eq.~\eqref{eq:VI}. The resulting equation for $\theta(0, t)$ is
\begin{align}\label{eq:shaps}
    &\frac{d\theta(0, t)}{dt}= \frac{d\theta^{(0)}(0, t)}{dt} + \frac{\pi}{2e} \frac{V_n}{R}\notag\\
    &\quad\quad-2\pi K J_n(\alpha)\lambda\sin\left(2\theta(0, t)- (\Omega - n \omega_{\rm ac})t\right).
\end{align}
It is further convenient to make an offset eliminating a constant term, $\theta(0, t) = \tilde{\theta}(0, t) + t(\pi / 2e) V_n / R$:
\begin{align}\label{eq:shaps_final}
    &\frac{d\tilde{\theta}(0, t)}{dt}= \frac{d\theta^{(0)}(0, t)}{dt}\\
    &-2\pi K J_n(\alpha)\lambda\sin\bigl(2\tilde{\theta}(0, t)- (\Omega - n \omega_{\rm ac} - \tfrac{\pi}{e} V_n / R)t\bigr).\notag
\end{align}
This equation reveals a special value of the DC bias, $I = 2e \Omega / 2\pi = n\cdot 2e \omega_{\rm ac} / 2\pi + V_n / R$, at which $d\langle\tilde{\theta}\rangle / dt = 0$. The latter condition means that the current through the resistor is $V_n / R$ and, accordingly, the voltage drop across it is $V_n$. The remaining part of the supplied current flows through the junction and is perfectly synchronized with the microwave radiation, $I_J = n\cdot 2e \omega_{\rm ac} / 2\pi$.
This defines the center of a step in the $V(I_J)$-relation.

\begin{figure}[t]
  \begin{center}\includegraphics[scale = 1]{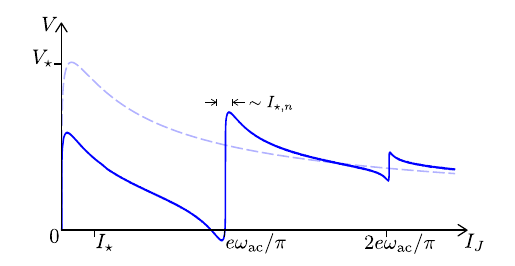}
    \caption{Bloch-Shapiro steps in the $V(I_J)$-relation of a transmon in {\bf circuit 1} [the curve is plotted with the help of Eqs.~\eqref{eq:shap_vi_1} and \eqref{eq:shap_vi_2} for $K = 1/8$].  The steps are centered at the values of current corresponding to the multiples of the microwave frequency, $I_J = n\cdot2e\omega_{\rm ac}/2\pi$. A finite width of the steps $I_{\star, n}$ [see Eq.~\eqref{eq:Istar_n}] stems from the zero-point fluctuations of charge transferred through the resistor. Dashed line shows $V(I_J)$ in the absence of microwaves.}
    \label{fig:shapiro}
  \end{center}
\end{figure}

To find the shape of a step, we note that the form of Eq.~\eqref{eq:shaps_final} is similar to that of an equation for $d\theta / dt$ in the DC case [cf.~Eq.~\eqref{eq:shaps0}].  Therefore, we can use the results of Sec.~\ref{sec:voltage_current} to find the voltage-current relation. With respect to its center, the step's shape is a replica of the DC $V(I_J)$. In the replica, the full current $I_J$ in the DC case is replaced by its deviation from the synchronized value, $I_J \rightarrow I_J - n \cdot 2e \omega_{\rm ac} / 2\pi$, while the voltage drop across the circuit is replaced by its deviation from the ``background'' value, $V\rightarrow V - V_n$. The phase slip amplitude is replaced by a value determined by the microwave amplitude, $\lambda \rightarrow J_n(\alpha) \lambda$.
In analogy with the DC case, we find that the qualitative character of $V(I_J)$ depends on the comparison between the deviation $I_J - n \cdot 2e \omega_{\rm ac} / 2\pi$ and current scale 
\begin{equation}\label{eq:Istar_n}
    I_{\star, n} = |J_n(\alpha)|^{1/(1-2K)}I_\star,
\end{equation}
where $I_\star$ is given by Eq.~\eqref{eq:I_star}, and $\alpha$ is the microwave amplitude cf.~Eq.~\eqref{eq:alpha}. This scale defines the width of the step. The step height is $V_{\star, n} \sim I_{\star, n} R$; its dependence on the microwave amplitude is $\propto |J_n(\alpha)|^{1/(1-2K)}$.
The tail of a step corresponds to $|I_J - n \cdot 2e \omega_{\rm ac} / 2\pi| \gg I_{\star, n}$ and is given by
\begin{equation}\label{eq:shap_vi_1}
V - V_n = \frac{\pi\lambda^2J^2_n(\alpha)}{2\Gamma(4K)}\frac{1}{I_J - n \tfrac{2e \omega_{\rm ac}}{2\pi}} \left(\frac{\pi|I_J - n \tfrac{2e \omega_{\rm ac}}{2\pi}|}{e\omega_{\rm Q}}\right)^{4K},    
\end{equation}
cf.~Eq.~\eqref{eq:VI}.
In the opposite limit, $|I_J - n \cdot 2e \omega_{\rm ac} / 2\pi| \ll I_{\star, n}$, the voltage changes sharply with the deviation from the step center:
\begin{align}\label{eq:shap_vi_2}
    V - V_n & = \mathrm{sign}\,\bigl(I_J - n \tfrac{2e\omega_{\rm ac}}{2\pi}\bigr)\notag\\
    &\times\frac{\omega_{\rm Q}}{2e} \left(\frac{\Gamma(1/K)\omega_{\rm Q}}{\pi \tilde{\lambda}_n^2} \frac{|I_J - n \tfrac{2e\omega_{\rm ac}}{2\pi}|}{e}\right)^{\tfrac{K}{1-K}}
\end{align}
[cf.~Eq.~\eqref{eq:V_vs_IJ_blockade}], where $\tilde{\lambda}_n$ is obtained from Eq.~\eqref{eq:tilde_lambda} by replacement $\lambda \rightarrow |J_n(\alpha)|\lambda$. The two asymptotes \eqref{eq:shap_vi_1} and \eqref{eq:shap_vi_2} match each other at $|I_J - n \cdot 2e \omega_{\rm ac} / 2\pi| \sim I_{\star, n}$. {Using the asymptotes, in Fig.~\ref{fig:shapiro} we sketch the $V(I_J)$-relation of the transmon in the presence of microwaves.}

The full shape of a step can be found in the limit of negligible quantum fluctuations, $R \gg R_Q$ (i.e., $K \ll 1$). In this case, the first term in the right hand side of Eq.~\eqref{eq:shaps_2} can be dropped. The solution of the resulting classical equation yields 
\begin{equation}
    V - V_n = (I_J - n\tfrac{2e\omega_{\rm ac}}{2\pi}) \left[\sqrt{1 + \left(\frac{4eKJ_n(\alpha)\lambda}{I_J - n\tfrac{2e\omega_{\rm ac}}{2\pi}}\right)^2} - 1\right],
\end{equation}
cf.~Eq.~\eqref{eq:classical_V_I_J}.
This formula shows that, in the presence of microwave radiation, the voltage-current relation develops a replica of the DC $V(I_J)$, with the magnitude of a jump rescaled by $|J_n(\alpha)|$. 
The dependence of the step height on the microwave amplitude  mirrors the conventional Shapiro steps \cite{tinkham2004}.

At finite $K$, the rescaling parameter changes to $|J_n(\alpha)|^{1/(1-2K)}$. The height of the steps rapidly decreases as $K$ approaches the Schmid transition point, $K \rightarrow 1/2$. There are no Bloch-Shapiro steps on the superconducting side of the transition, $K > 1/2$.

{Synchronization between the Bloch oscillations and the microwave radiation can also be achieved in the circuit of Fig.~\ref{fig:circuit2}. It would lead to steps in 
$V(I)$ centered around $I = n\cdot 2e \omega_{\rm ac} / 2\pi$, with a similar dependence of step width and height on the  microwave amplitude $\alpha$ and impedance~$R$.}



\section{Discussion and Conclusions}
The Coulomb blockade impedes the flow of supercurrent through the Josephson junction embedded into a high impedance environment (formed by, e.g., a resistor or a junction array). Depending on the environment, the junction's ground state may change from a superconducting to an insulating one. 
In the superconducting state, the differential resistance $dV/dI_J$ of the junction
approaches zero at $I_J \rightarrow 0$, while $dV / dI_J$ of an insulating junction diverges at low current due to the onset of the Coulomb blockade.
The transition between the superconducting and insulating ground states occurs at the critical value of the impedance $R = R_Q = h / (2e)^2$ regardless of the relation between the charging energy $E_C$ and the Josephson energy $E_J$. 

However, the relation between $E_C$ and $E_J$ determines how big is the portion of $VI$-characteristic where the junction exhibits the insulating behavior. Specifically, the insulating state of a transmon circuit ($E_J\gg E_C$) is confined to a domain $I_J \lesssim I_\star$ exponentially small in $E_J / E_C \gg 1$, see Eqs.~\eqref{eq:phase_slip} and \eqref{eq:I_star}. At higher currents, the circuit $VI$-characteristic is hardly distinguishable from that of a 
conventional superconducting junction, see, e.g., Fig.~\ref{fig:circuit1}(b). Nonetheless, the Cooper pair charge discreteness, which gives rise to the Coulomb blockade in the first place, continues to manifest at $I_J \gtrsim I_\star$. The most striking manifestation is the Bloch oscillations of voltage across the Josephson junction. Despite exponential smallness of $I_\star$, transmon is the most suitable circuit for the observation of this phenomenon. The reason is a large gap between the lowest-energy charge band and higher-energy  states, preventing the transmon from ionization by inter-band transitions.

Classically, the Bloch oscillations occur at frequency $\Omega_J = 2\pi I_J / 2e$ set by the current flowing through the junction and the Cooper pair charge $2e$. Quantum fluctuations broaden the oscillations spectrum; the monochromatic line at $\Omega_J$ is replaced by a power-law threshold behavior $\propto \Theta(\Omega_J - \omega)$, see Eq.~\eqref{eq:S_omega}. The broadening of the spectrum increases with the impedance $R$ being lowered towards~$R_{Q}$. 

Bloch oscillations can be synchronized with the externally applied microwave radiation. Synchronization results in the formation of steps in the $VI$-characteristic of the circuit. These steps are dual to the conventional Shapiro steps, and occur at the quantized values of current $I_J = n \cdot 2e \omega_{\rm ac} / 2\pi$ set by the multiples of microwave frequency $\omega_{\rm ac}$. The height of the $n$-th step, $V_{\star , n} \propto |J_n(\alpha)|^{1 / (1 - R_Q / R)}$, depends on the impedance $R$ as well as on the microwave amplitude $\alpha$. Due to the quantum fluctuations of charge, the steps are not perfectly vertical even at zero temperature.  The step width is $I_{\star , n} \sim V_{\star, n} / R$, see Eq.~\eqref{eq:Istar_n}.  
The specific shape of the steps is given by Eqs.~\eqref{eq:shap_vi_1} and \eqref{eq:shap_vi_2}. 
In the limit $R \rightarrow \infty$, the effect of quantum fluctuations becomes negligible. Then, Bloch-Shapiro steps are an exact dual of classical Shapiro steps whose height scales with the first power of the Bessel function and which are perfectly sharp \cite{tinkham2004}. 
The dual Shapiro steps vanish across the Schmid transition, $R = R_Q$, along with other effects of charge discreteness. 

There is a duality between the charge and phase fluctuations. The latter establish a fundamental limitation for the sharpness of the conventional Shapiro steps. Upon the proper change of variables $V\leftrightarrow I$ and impedances $R\leftrightarrow R_Q^2/R$, the quantum-broadened Shapiro and Bloch-Shapiro steps are described by the same dimensionless function. One may ask a question: how high should $R$ be for the Bloch-Shapiro steps to be as sharp as the conventional Shapiro steps for a Josephson junction coupled to vacuum impedance of $370\,\Omega$? The answer is $R\approx 110\,{\rm k\Omega}$. A comparable impedance was used in the experiment \cite{shaikhaidarov2022}, but the width of the observed Bloch-Shapiro steps exceeded substantially the quantum limit. This means that other mechanisms (e.g., the microwave-induced heating) are responsible for the steps width; elimination of such mechanisms can significantly improve the accuracy of current quantization. 

\section{Acknowledgements}
This work was supported by
NSF Grant No. DMR-2002275 and by ARO Grant No.
W911NF-23-1-0051. BR acknowledges the support of Yale Prize Postdoctoral Fellowship in Condensed Matter Theory.

\bibliography{references.bib}

\end{document}